\documentstyle[preprint,prl,aps]{revtex}
\begin{document}
{\bf 
\centerline {A quasi-random spanning tree model for the early river network}
}
\vskip 1.0 cm
\centerline {S. S. Manna$^{1,2,}$\footnote{manna@niharika.phy.iitb.ernet.in} and B. Subramanian$^2$}
\vskip 1.0 cm
\begin {center}
$^{1}$H\"ochstleistungsrechenzentrum des Forschungszentrum,
D-52425 J\"ulich , Germany \\
$^{2}$Department of Physics,
Indian Institute of Technology,
Bombay 400076, India \\
\end {center}

{\bf Abstract :} We consider a model for the formation of a 
river network in which erosion process plays a role only at the
initial stage. Once a global connectivity is achieved, no
further evolution takes place. In spite of this, the
network reproduces approximately most of the empirical
statistical results of natural river network. It is observed
that the resulting network is a spanning tree graph
and therefore this process could be looked upon as a new algorithm
for the generation of spanning tree graphs in which 
different configurations occur quasi-randomly.
A new loop-less percolation model is also defined at an intermediate
stage of evolution of the river network.
\eject
\vfill
A river network consists of a main river accompanied by
a hierarchy of side streams of decreasing lengths and 
flow capacities. Ignoring the ground absorption
and evaporation, the network drains out the whole amount of
rain water dropped uniformly on every 
small piece of land in the river basin
and therefore necessarily spans the whole drainage area.
In addition, though it is quite common that 
two rivers join together, it is hardly 
observed that a river bifurcates into two smaller streams 
(ignoring small delta islands)
because of the fact that water flows in the direction of steepest descent.
Therefore the general structure of a river network is like
a tree on which two points are connected by a distinct path.
Due to these properties, a river network qualifies
to be described by a spanning tree graph, 
the loop-less graph which covers all nodes,
a well known example in graph theory [1].
The aim of this paper is to propose a quasi-random
spanning tree model for the formation of a river network 
from the very early stage.

The geometrical structure of the network considering
different streams as linear segments has  been 
of considerable interest for a long time.
Quite commonly different rivers are classified according to Strahler's
ordering procedure [2]. In this recursive ordering scheme,
two streams of orders $n_1$ and $n_2$ meet to produce
a stream of order $n$ as :
\begin{equation}
{\rm if}\quad n_1 \ne n_2 \quad{\rm then}\quad n = \max(n_1,n_2);\quad
        {\rm else,}\quad n=n_1+1
\end{equation}
where the streams which start from sources are assigned order $n=1$.
Horton empirically observed
that the average number and length of rivers
of different orders in a network follow geometric
series with approximately constant bifurcation $r_b$ and
the length ratio $r_l$ [3]. Mandelbrot suggested that
the river network might be a self-similar fractal with fractal
dimension $d_{rn}=2$ because of the spanning nature of the network [4].
Later, fractal dimension of the individual rivers $d_c$ are related
to the Horton's ratios by [5,6]
\begin{equation}
d_{rn} = d_c \frac{\log r_b}{\log r_l}.
\end{equation}
The mean annual discharge at any link or its surrogate variable,
the cumulative area contributing to the link
follows a power law probability distribution
as, $P(a) \sim a^{-\tau_a}$ [7].
The length $l$ of a typical stream of certain order
also follows a power law distribution
$P(l) \sim l^{-\tau_l}$ [8].
Another empirical result is that the average length 
$<l_\Omega>$ of the river with maximum order $\Omega$
varies with the basin area $a$ as $l_\Omega \sim a^\alpha$ [9].

Scheidegger proposed a lattice model of directed river network 
defined on a slope [10]. This model was shown to be identical
to the one dimensional random particle aggregation model [11].
More recently, following the idea of Self-Organized Criticality
[12], a number of river network models are proposed
which successfully produce spatial scale invariance in 
the self-organized critical states [13].

The problem of spanning tree graphs is well known in
statistical physics.
Kirchhoff related the spanning tree graphs to
the problem of
determining the effective resistance between two nodes
of a resistor network [14]. Fortuin and Kasteleyn showed
that it is related to the $q \rightarrow 0$ limit of the $q$-state
Potts model [15]. Recently the correspondence between the
spanning tree graphs and the steady state configurations
of the Abelian sandpile model in the Self-Organized Criticality [16]
is established.
In the case of random spanning tree problem
all possible tree configurations occur with equal
probability and this model is very well studied. We 
compare the results of our model with those of random spanning trees and
conclude that our model belongs to a new universality class.

The process of erosion is the underlying mechanism for the evolution of
a river network. Erosion takes place during the flow of streams
which modifies the river beds and therefore causes changes in the
flow pattern.
We consider here the evolution of a network 
to its full connected form starting from the very initial
stage of isolated lakes. 

Continuously variable heights with uniform random distributions
are assigned to all sites of a square lattice. We first assume that
rain falls only along the bonds of the lattice
and flows downwards on the slope along the bond.
Due to this unidirectional flow, intensive erosion 
process takes place which reduces 
the slope along the bond. We assume that this reduction 
of slope due to the erosion is a very slow process 
since huge amounts of sediments are transported from 
one place to the other. When the slope decreases
water gets accumulated in different places
along the bond and finally when the slope is very small
a little further rain fall in a very short time 
floods the whole length of the bond,
forming a lake of the smallest size of only one bond. We assume that this
transition time is much smaller than the time required for the
whole erosion process.

Rain falls simultaneously along all the
bonds of the lattice, the erosion process takes 
place in parallel and lattice bonds
become lakes one after another sequentially
occuring with uniform probability.
Several one bond lakes join together and form bigger lakes.     
We assume that all sites
of one lake have approximately equal heights and since
the water in a lake is stagnant, there is no significant
erosion takes place to change their heights further. 

For the case of a bond which does not belong to any lake
but the adjacent two sites belong to the same lake, situation 
is different. Since the sites are approximately of same height,
there is no significant height gradient along the bond
and the erosion process will not be  fast enough
to equalize the level of the bond with the sites. 
Therefore this bond will never be included
into the lake forbidding the possiblity of loop formation.

The role of a lake is to store rain water in the
initial stage, untill it gets connected to a flowing
river when it also starts flowing and 
becomes part of the river network.
The first site on the 
boundary of the lattice which becomes the member of a particular lake
is connected to the ocean outside.
This creates a net directed flow in every bond of this lake
which form a small cluster of rivers.
In a similar manner all other lakes also eventually become
connected among themselves and to this small cluster of rivers
and therefore will start flowing. 
Positive slopes against the 
flow are created due to erosion along all the bonds 
which completes the formation of the river network.

A random list of all bonds $B$ of the lattice is
generated from an ordered sequence by large number of random
pair interchanges. Here the computational effort increases
linearly with $B$ to obtain a most uncorrelated configuration,
compared to $B \log B$ in the Broder's algorithm for generating
the random spanning trees [17].

Numbers are called sequentially from this random list
and corresponding bonds on the lattice are tested for the
lake formation. A bond is allowed to be a lake if it is the 
smallest lake, becomes part of the bigger lake or
joins two distinct lakes. A bond is forbidden to be 
occupied by a lake if it connects two sites of the same lake.
We use the Hoshen and Kopelman's algorithm for cluster 
numbering in percolation theory [18] 
for identifying different lakes and to restrict
loop formations. Finally a single connected network in the form of a
spanning tree graph covers the whole lattice (Fig. 1).

We first exactly calculate the probabilities of occuring for
different spanning trees generated from 
random permutations on a $2 \times 3$ cell and
observe that they are non-uniformly distributed (Fig. 2).
We conclude that spanning trees obtained from random 
bond permutations occur with non-uniform 
probabilities and therefore we call them as `quasi-random'. 

Periodic boundary conditions are used in all four directions
of the square lattice
and the outlet of the network is chosen
at the $root$ of the spanning tree.
We first study the connectivity of a randomly chosen site.
The average fractions of sites connected to 1, 2, 3 and
4 bonds are obtained as 0.30681, 0.42698, 0.22557 and 0.04061,
slightly different from their counterparts 
0.29454, 0.44699, 0.22239 and 0.03608
in random spanning trees [19].

The drainage area at the site $i$ is defined as 
$
a_i = \Sigma_j w_{ij}a_j +1 
$
where $j$ runs over the nearest neighbour sites and $w_{ij}$ = 1
if flow direction is from $j$ to $i$, otherwise it is zero.
Area values are calculated using a systematic deleting procedure.
Leaf sites of the network are the set of sites 
connected by only one bond, initially unit area values
are assigned to them.
At the deleting time $t$ all the leaf sites are deleted simultaneously
and area values are carried over to the connected sites.
This creates a new set of leaf sites to be deleted in the
time $t+1$. The root gets an area $L^2$.
In the figure 3 we plot the probability
distribution of the drainage area $P(a)$ for $L=1024$
and obtain a very nice straight line.
We estimate $\tau_a = 1.392 \pm 0.010$ and compare with 11/8 for random
spanning trees [19] and to 4/3 of the directed river network model [10,11].
Empirical values of $\tau_a $ varies from
1.41 to 1.44 for different river basins [7].

To calculate the stream length distribution $P(l)$
we delete different streams 
sequentially one after the other.
Deletion starts from a leaf site,
proceeds along the river and stops when the river
meets a higher order river. When all first order rivers are
deleted, we get another set of leaf sites all of which correspond to
the second order rivers, which are also eventually deleted.
Using $L=1024$ we get
the exponent $\tau_l = 2.65 \pm 0.03$, 
to be compared with its empirical value 2.9 [8].

The average length $<l_\Omega>$ of the rivers with
maximum order number $\Omega$ varies with the 
whole basin area $L^2$ with a power $\alpha = 0.636 \pm 0.005$.
Empirically one gets $\alpha = 0.58 \pm 0.03$ [9].

We also studied the statistics of the longest river
flowing into a site. The deleting time of any site is the length
of the longest river $l_m$ at that site.
This length also follows a power law distribution 
$P(l_m) \sim l_m^{-\tau_m}$ with $\tau_m = 1.628 \pm 0.005$.
Similarly defined exponent $\alpha_m$ in $l_m \sim <a>^{\alpha_m}$
is obtained as $0.608 \pm 0.005$.
This gives a connection between $a$ and $l_m$ as
$P(a)da = P(l_m)dl_m$ and the scaling relation 
$\alpha_m =(\tau_a-1)/(\tau_m-1)$ gives approximately the
same value of $\alpha_m$ as measured numerically.

The chemical distance between any two points is defined as the
length of the shortest connecting path.
Therefore to calculate the fractal dimension of the rivers in our model
we calculate the dimension of the chemical paths.
A reverse deleting of the network is done from the root of the
tree. The deleting time of a site is the length of the river
to that site from the root. The probability that an arbitrarily
selected site is at a chemical distance $l_r$ from
the root follows a scaling form $P(l_r,L)=L^{d_c}f(l_r/L^{d_c})$.
From an excellent data collapse of this distribution data for 
$L$ = 64, 256 and 1024 we get $d_c = 1.217$.
The scaling function also fits very well to the form
$f(x) = ax^bexp(-cx^d)$ where $a = 1.30, b = 0.59, c = 0.60, d = 2.62$.
We also get another value of $d_c =1.222$ by directly
calculating the average length of the river $<l_r(L)> \sim L^{d_c}$.
We conclude $d_c = 1.220 \pm 0.010$ and
compare it with the random spanning tree value 5/4 [20].

In figure 4 we plot the average number of rivers $<N_n>$ and the 
average length of the rivers $<l_n>$ for different orders $n$ for a lattice
of length $L=1024$. We obtain almost constant value of 
Horton's bifurcation ratio $r_b = 4.39$ and the length ratio $r_l = 2.44$
in the region from $n$ = 2 to 7.
We compare these values with Shreve's calculation
of $r_b= 4 $ and $r_l=2$ for the equally weighted river networks [21].
Using the eqn. (2) and the fractal dimension of the
rivers $d_c =1.220$ we calculate 
that fractal dimension of the river network $d_{rc}$ = 2.02
which is quite close to its exact value 2 for our quasi-random
spanning tree river network.

Finally we consider the situation where rain falls 
non uniformly or the basin area contains some 
randomly positioned dry lands using 
a new percolation model.
We randomly throw bonds on the lattice 
in the same way as before but keep
checking if the connectivity is formed
between any two opposite sites of the lattice. The moment
it is formed we stop further dropping of bonds.
We see that $p_c(\infty)-p_c(L) \sim L^{-1/\nu}$ where
$p_c(\infty) = 0.4511 \pm 0.0005$ and $\nu = 1.334 \pm 0.005$,
which is very close to the value of $\nu = 4/3$ but the value
of $d_c = 1.119 \pm 0.005$ is distinctly different from 
$1.1307 \pm 0.0004$ for ordinary percolation [22].
Percolation on the Bethe lattices are previously considered by
Straley [23].

To summarize, we have considered the formation of a globally connected river
network starting from the very early stage of water
accumulation in the microscopic lakes. Lakes grow in size
and eventually get connected to the ocean when its different branches
become rivers. Finally the river network
spans the basin. We model the river network by a quasi-random
spanning tree belonging to a new universality class. 
We see that though we have not considered the
temporal development of rivers, the first connected network
closely reproduces the statistics of the natural river network.
We also study a new loop-less perolation model at an intermediate 
stage of evolution of the river network.

After finishing this work we came to know about the work
of Cieplak et. al. [24] who considered disorder-dominated
river basins and obtained results similar to us.

We thank D. Dhar for many useful suggestions,
H. Kallabis for much help in graphics
and A. Giacometti and D. Wolf for the
critical reading of the manuscript.
\eject
\vfill
\vskip 0.5 cm
{\bf Figure Captions}
\vskip 0.5 cm
Figure 1 : A typical quasi-random spanning tree configuration
for modelling the river network
on the $32 \times 32$ lattice. Rivers of order 1(black), 2(yellow),
3(blue), 4(green) and 5(red) are shown. Connection to the ocean
is through the site with a circle at the bottom.
\vskip 0.5 cm
Figure 2 : The fifteen distinct spanning tree configurations
on a $2 \times 3$ lattice. The nine type (a) configurations
occur 360 times and the six type (b) configurations occur
300 times in all the spanning tree configurations
generated by exact enumeration of the $7!=5040$ permutations
of the seven bonds of the lattice.
\vskip 0.5 cm
Figure 3 : The probability distribution 
$P(a)$ of finding an arbitrarily
selected site of drainage area $a$ is shown for $L$ =1024,
which gives $\tau_a = 1.392$.
\vskip 0.5 cm
Figure 4 : The average number of rivers $<N_n>$ (denoted by circles)
and the average length of the rivers $<l_n>$ (denoted by crosses) 
for different order numbers are plotted against $n$.
The bifurcation ratio $r_b = <N_n>/<N_{n+1}>$
and length ratio $r_l = <l_{n+1}>/<l_n>$
are obtained 4.39 and 2.44 respectively.

\eject
\vfill
\vskip 0.5 cm
{\bf References}
\vskip 0.5 cm
[1] F. Harary, {\it Graph Theory} (Addition-Wesley, Reading, MA, 1990).

[2] A. N. Strahler, Trans. Am. Geophys. Union {\bf 38}, 913 (1957).

[3] R. E. Horton, Geol. Soc. Am. Bull. {\bf 56}, 275 (1945).

[4] B. B. Mandelbrot, {\it The Fractal Geometry of Nature}, W.
H. Freeman, New York (1983).

[5] P. La Barbera and R. Rosso, Water Resour. Res. {\bf 25}, 735 (1989).

[6] D. G. Tarboton, R. L. Bras and I. Rodriguez-Iturbe, Water Resour. Res.
{\bf 26}, 2243 (1990).

[7] I. Rodriguez-Iturbe, E. J. Ijj\`asz-V\`asquez, 
R. L. Bras and D. G. Tarbotton,
Water Resour. Res. {\bf 28}, 1089 (1992).

[8] D. G. Tarboton, R. L. Bras and I. Rodrigues-Iturbe, Water Resour. Res.,
   {\bf 24}, 1317 (1988).

[9] J. T. Hack, U. S. Geol. Surv. Prof. Pap. {\bf 294-B} (1957).

[10] A. E. Scheidegger, Bull. I.A.S.H. {\bf 12} (1), 15 (1967).

[11] H. Takayasu, Phys. Rev. Lett. {\bf 63}, 2563 (1989);
T. Nagatani, J. Phys. A. {\bf 26}, L489 (1993).

[12] P. Bak, C. Tang and K. Wiesenfeld, Phys. Rev. lett. {\bf 59},
381 (1987).

[13] H. Takayasu and H. Inaoka, Phys. Rev. Lett. {\bf 68}, 966 (1992);
 A. Rinaldo, I. Rodriguez-Iturbe, R. Rigon, E. Ijjasz-Vasquez 
and R. L. Bras, Phys. Rev. Lett. {\bf 70}, 822 (1993);
A. Giacometti, A. Maritan and J. R. Banavar, Phys. Rev. Lett. {\bf 75},
577 (1995).

[14] G. Kirchhoff, Ann. Phys. Chem. {\bf 72}, 497 (1847).

[15] C. M. Fortuin and P. W. Kasteleyn, Physica {\bf 57}, 536 (1972).

[16] S. N. Majumdar and D. Dhar, Physica A {\bf 185}, 129 (1992).

[17] A. Z. Broder, in {\it Proceedings of the 30-th Annual IEEE
Symposium on Foundations of Computer Science}, 
(IEEE, New York, 1989), p.442.

[18] J. Hoshen and R. Kopelman, Phys. Rev. B., {\bf 14}, 3428 (1976)

[19] S. S. Manna, D. Dhar and S. N. Majumdar, Phys. Rev. A., 
{\bf 46}, 4471 (1992).

[20] B. Duplantier, J. Stat. Phys. {\bf 54}, 581 (1989); A. Coniglio,
Phys. Rev. Lett. {\bf 62}, 3054 (1989).

[21] R. L. Shreve, J. Geol. {\bf 74}, 17 (1966); {\bf 77}, 397 (1969).

[22] P. Grassberger, J. Phys. A. {\bf 25}, 5475 (1992).

[23] J. P. Straley, J. Phys. C. {\bf 9}, 783 (1976).

[24] M. Cieplac, A. Giacometti, A. Maritan, A. Rinaldo, 
I Rodriguez-Iturbe and J. R. Banavar, unpublished.

\end{document}